\documentclass[twocolumn,english,aps,pra,superscriptaddress]{revtex4}
\usepackage[LGR,LGR,LGR,LGR,LGR,T1]{fontenc}
\usepackage[latin9]{inputenc}
\usepackage{color}
\usepackage{units}
\usepackage{amsmath}
\usepackage{amssymb}
\usepackage{graphicx}

\makeatletter

\DeclareRobustCommand{\greektext}{%
  \fontencoding{LGR}\selectfont\def\encodingdefault{LGR}}
\DeclareRobustCommand{\textgreek}[1]{\leavevmode{\greektext #1}}
\ProvideTextCommand{\~}{LGR}[1]{\char126#1}

\providecommand{\tabularnewline}{\\}

\@ifundefined{textcolor}{}
{%
 \definecolor{BLACK}{gray}{0}
 \definecolor{WHITE}{gray}{1}
 \definecolor{RED}{rgb}{1,0,0}
 \definecolor{GREEN}{rgb}{0,1,0}
 \definecolor{BLUE}{rgb}{0,0,1}
 \definecolor{CYAN}{cmyk}{1,0,0,0}
 \definecolor{MAGENTA}{cmyk}{0,1,0,0}
 \definecolor{YELLOW}{cmyk}{0,0,1,0}
}

\usepackage{babel}
\usepackage[normalem]{ulem}

\usepackage{babel}

\usepackage{babel}

\usepackage{babel}

\makeatother

\usepackage{babel}
\begin{document}

\title{Mutually Unbiased Coarse-Grained Measurements of Two or More Phase-Space
Variables}

\author{E. C. Paul}
\email{eduardo.paul@if.ufrj.br}

\selectlanguage{english}%

\affiliation{Instituto de F{í}sica, Universidade Federal do Rio de Janeiro,
Caixa Postal 68528, Rio de Janeiro, RJ 21941-972, Brazil}

\author{S. P. Walborn}

\affiliation{Instituto de F{í}sica, Universidade Federal do Rio de Janeiro,
Caixa Postal 68528, Rio de Janeiro, RJ 21941-972, Brazil}

\author{D. S. Tasca}

\affiliation{Instituto de F{í}sica, Universidade Federal Fluminense, Niteroi,
RJ 24210-346, Brazil}

\author{\L ukasz Rudnicki}
\email{rudnicki@cft.edu.pl}

\selectlanguage{english}%

\affiliation{Max-Planck-Institut f{ü}r die Physik des Lichts, Staudtstra{ß}e
2, 91058 Erlangen, Germany}

\affiliation{Center for Theoretical Physics, Polish Academy of Sciences, Aleja
Lotnik{ó}w 32/46, 02-668 Warsaw, Poland}
\begin{abstract}
Mutual unbiasedness of the eigenstates of phase-space operators\textendash such
as position a\textcolor{black}{nd momentum, or their standard coarse
grained versions\textendash exists only in the limiting case of infinite
squeezing. In {[}Phys. Rev. Lett. }\textbf{\textcolor{black}{120}}\textcolor{black}{,
040403 (2018){]} it was shown that mutual unbiasedness can be recovered
for periodic coarse grainings of these two operators. Here we investigate
mutual unbiasedness of coarse-grained measurements for more than two
phase-space variables. We show that mutual unbiasedness can be recovered
between periodic coarse graining of any two non-parallel phase-space
operators. We illustrate these results through optics experiments,
using the fractional Fourier transform to prepare and measure mutually
unbiased phase-space variables. The differences between two and three
mutually unbiased measurements is discussed. Our results contribute
to bridging the gap between continuous and discrete quantum mechanics
and could be useful in quantum information protocols. }
\end{abstract}
\maketitle

\section{Introduction}

Complementarity\textendash the fact that \textcolor{black}{perfect}
knowledge about a certain observable prohibits the knowledge of a
second complementary observable\textendash is a cornerstone of Quantum
Mechanics and Quantum Information. Complementarity can be better formulated
through the concept of mutual unbiasedness \cite{Schwinger60}, which
can be characterized in terms of bases (projectors) or more generalized
measurements \cite{Kalev14}. Mutually unbiased bases (MUBs) \cite{Kraus87}
play an important role in the security of quantum cryptography \cite{Gisin02,Coles17},
the efficiency of quantum tomography \cite{Fernandez-Perez11,Giovannini13},
and are useful for identifying quantum correlations such as entanglement
\cite{Spengler12,Paul16,Sauerwein17} and steering \cite{Cavalcanti09,Walborn11,Tasca13a,Schneeloch13a,Schneeloch13b,Skrzypczyk14,Marciniak15,Zhu16},
as well as for certifying quantum randomness \cite{Vallone14}.

In a finite $d$-dimensional Hilbert space, two orthonormal bases
$\{\left|a_{i}\right\rangle \}$ and $\{\left|b_{j}\right\rangle \}$
are mutually unbiased if and only if $\left|\left\langle a_{i}\left|b_{j}\right\rangle \right.\!\right|=1/\sqrt{d}$,
for $i,j=0,...,d-1$ \cite{Durt10}. For the case of continuous variables
(CV), it is well known that the conjugate pair of position and momentum
eigenstates also presents mutual unbiasedness, as per the relation
$\left|\left\langle x\left|p\right\rangle \right.\!\right|=1/\sqrt{2\pi}$,
where here and throughout we set $\hbar=1$. The kets $\left|x\right\rangle $
and $\left|p\right\rangle $ are normalized (to Dirac delta) eigenvectors
of $\hat{x}$ (position) and $\hat{p}$ (momentum) operators, respectively.
In fact, if we define rotated phase-space operators as linear combinations
of \textit{dimensionless} position and momentum operators, 
\begin{equation}
\hat{q}_{\theta}=\cos\theta\hat{x}+\sin\theta\hat{p},\label{q_theta}
\end{equation}
the eigenbases of $\hat{q}_{\theta}$ and $\hat{q}_{\theta'}$ are
always mutually unbiased, provided that $\sin\left(\theta-\theta'\right)\neq0$.

\begin{figure}
\centering{}\includegraphics{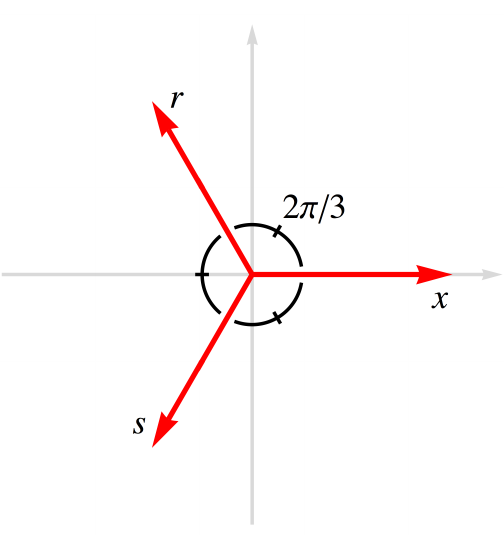}\caption{\label{fig:3-bases}Pictorical representation of the phase-space variables
defining the 3 mutually unbiased bases in continuous variables. The
arrows divide the phase space in three equal slices of $2\pi/3$.}
\end{figure}

While mutual unbiasedness is not limited to only two phase-space operators
in the CV case, there is a large qualitative difference when compared
to discrete, finite-dimensional quantum mechanics. In particular,
for a $d$-dimensional system, the condition for MUBs can be mutually
satisfied by $d+1$ bases whenever $d$ is the power of a prime number
\cite{Durt10}. For CVs, even though one might suspect that the infinite-dimensional
nature of the state space would lead to infinitely many MUBs, this
is not the case. Indeed, Weigert and Wilkinson \cite{Weigert08} have
shown that one can identify at most three mutually unbiased bases
for each CV system. This result applies to the bases mutually related
by rotations on the phase space, leading to a mutually unbiased phase-space
\textquotedbl{}triple\textquotedbl{} set of operators. For example,
in addition to the eigenbasis of the position operator $\hat{x}\equiv\hat{q}_{0}$,
one can consider eigenbases of two more operators: $\hat{r}\equiv\hat{q}_{2\pi/3}$
and $\hat{s}\equiv\hat{q}_{4\pi/3}$, as illustrated in Fig.~\ref{fig:3-bases}.
The choice to begin with the position operator has been made without
loss of generality \textemdash{} alternatively one could start with
any other $\hat{q}_{\theta}$, and rotate $\hat{r}$ and $\hat{s}$
by $\theta$. Furthermore, if squeezing operations are allowed, mutual
unbiasedness can be preserved even for phase space triples whose angles
do not differ by $2\pi/3$ \cite{Weigert08}.

The above discussion on MUBs in the context of CVs was based on the
definition of eigen\textcolor{black}{states of phase-space operators.
These states are normalized to Dirac delta distributions and as such
are not accessible in an experiment, as they correspond to the regime
of infinite squeezing \cite{Braunstein05}. When considering the physical
case of quantum states that are merely \textquotedbl{}localized\textquotedbl{}
around some\textendash say\textendash position, the corresponding
MUB condition with the momentum is lost. This is true for all mutually
unbiased pairs as well as for triples.}

\textcolor{black}{A related issue arises in experimental scenarios
where it is beneficial or even necessary to use coarse-grained measurements
\cite{Rudnicki12a,Rudnicki12b,Schneeloch13a,Tasca13a,Tasca17a}. In
\cite{Tasca17a}, while working with the pair $\left(x,p\right)$
of canonically conjugate variables, it was shown that ``standard''
coarse graining does not preserve the unbiasedness property originally
present in CVs. Instead, one can work with coarse-grained periodic
structures and recover unbiasedness in that regime, in some way opposite
to that known from modular variables \cite{Aharonov69,Busch86,Reiter89,Gneiting11,Carvalho12,Vernaz-Gris14,Ketterer16,Laversanne-Finot15}.
The goal of the present paper is to expand the theoretical and experimental
analysis performed in Ref. \cite{Tasca17a} to the case of general
phase-space variables, and in particular, to coarse-grained measurements
of more than two phase-space operators.}

\textcolor{black}{This paper is organized in the following way. In
section \ref{sec:pcg} we introduce periodic coarse-grained measurements
along arbitrary directions in phase space. In section \ref{sec:MUB2}
we define mutual unbiasedness of pairs of coarse-grained operators,
and extend this definition to more than two measurements in section
\ref{sec:MUB3}. In section \ref{sec:exp} we present experiments
and results investigating both the mutual unbiasedness of PCG phase-space
triples, and the mutual unbiasedness of PCG measurements corresponding
to phase-space directions related by an arbitrary angle. Finally,
we provide concluding remarks in section \ref{sec:conc}.}

\section{Periodic Coarse-graining of phase-space variables}

\label{sec:pcg}

Let $\left|q_{\theta}\right\rangle $ denote an eigenstate of the
quadrature operator $\hat{q}_{\theta}$ defined in Eq. \eqref{q_theta}.
The scalar product between eigenstates corresponding to different
quadrature operators is $\left\langle q_{\theta'}\left|q_{\theta}\right\rangle \right.\!=\mathcal{F}\left(q_{\theta'},q_{\theta}\right)$,
where 
\begin{equation}
\mathcal{F}\left(q_{\theta'},q_{\theta}\right)=\sqrt{\frac{ie^{i\Delta\theta}}{2\pi\left|\sin\Delta\theta\right|}}e^{i\frac{\cot\Delta\theta}{2}\left(q_{\theta}^{2}+q_{\theta'}^{2}\right)-i\frac{q_{\theta}q_{\theta'}}{\sin\Delta\theta}},
\end{equation}
is the kernel of the fractional Fourier transform \cite{Ozaktas01},
and $\Delta\theta=\theta'-\theta$. As $\left|\left\langle q_{\theta'}\left|q_{\theta}\right\rangle \right.\!\right|=1/\sqrt{2\pi\left|\sin\Delta\theta\right|}$,
it is clear that the eigenbases of any two operators $\hat{q}_{\theta}$
and $\hat{q}_{\theta'}$ are mutually unbiased whenever $\sin\Delta\theta\neq0$.

We can use the eigenstates of $\hat{q}_{\theta}$ to define a family
(labeled by $\theta$) of $d$ coarse-grained projective measurement
operators 
\begin{equation}
\hat{\Omega}_{k}^{\theta}=\int dq_{\theta}M_{k}\left(q_{\theta}-q_{\theta}^{\textrm{cen}};T_{\theta}\right)\left|q_{\theta}\right\rangle \left\langle q_{\theta}\right|,\label{Projectors}
\end{equation}
with detector apertures encoded in ``mask functions'' $M_{k}$,
such that \textcolor{black}{$\sum_{k=0}^{d-1}M_{k}=1$. The displacement
parameter $q_{\theta}^{\textrm{cen}}$ is included to represent the
freedom of setting the origin, and the parameter $T_{\theta}$ is
the period of the mask function. One possible choice for the mask
functions is 
\begin{equation}
M_{k}\left(z;T\right)=\begin{cases}
1, & ks\leq z\;\left(\textrm{mod }T\right)\leq\left(k+1\right)s\\
0, & \textrm{otherwise}
\end{cases}.\label{eq:mask-definition}
\end{equation}
}

\begin{figure}
\includegraphics{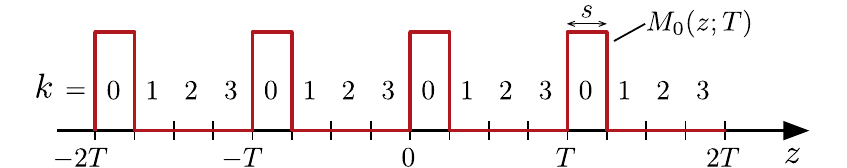}

\caption{\label{fig:Mask}Mask function $M_{0}\left(z;T\right)$, as defined
in Eq. \ref{eq:mask-definition}, for the $d=4$ case. Here, $T$
is the mask's spatial period, $s$ is its bin width, $z$ is the spatial
coordinate along which the mask is defined, and $k$ is the index
that identifies individual masks. This mask can be viewed as a periodic
array of apertures spaced by $d=4$ bins of size $s$.}
\end{figure}

\textcolor{black}{As defined in Ref. \cite{Tasca17a}, the functions
\eqref{eq:mask-definition} are periodic square waves specified by
the period $T$ and bin width $s=T/d$, so that $d$ can be considered
as a \textquotedbl{}dimensionality\textquotedbl{} parameter. The outcome
probabilities produced by the set of projectors \eqref{Projectors}
with the periodic mask functions \eqref{eq:mask-definition} define
the PCG of the probability distribution associated with the phase-space
variable $q_{\theta}$. Since we work with dimensionless variables,
both mask parameters $T$ and $s$ are also }dimensionless. In Fig.
\ref{fig:Mask} we illustrate the PCG geometry and the periodic mask
function for the particular case of $d=4$.

The operators $\hat{\Omega}_{k}^{\theta}$ are diagonal in their associated
\textcolor{black}{basis} given by $\{\left|q_{\theta}\right\rangle \}$.
In order to discuss mutual unbiasedness between several PCG measurements
corresponding to different directions in phase space, it is beneficial
to express $\hat{\Omega}_{k}^{\theta}$ in terms of an arbitrarily
rotated basis $\left|q_{\theta'}\right\rangle $. To this end, one
needs to represent the periodic mask function by means of its Fourier
series decomposition 

\begin{subequations}\label{FourierSeries} 
\begin{equation}
M_{k}\left(z;T\right)=\sum_{N\in\mathbb{Z}}f_{N}e^{-\frac{2\pi iN}{d}k}e^{\frac{2\pi iN}{T}z},
\end{equation}
with 
\begin{equation}
f_{N}=\frac{1-e^{-\frac{2\pi iN}{d}}}{2\pi iN}.\label{prefactor}
\end{equation}
\end{subequations} 

A moderately straightforward but lengthy calculation (see Appendix
\ref{ApA}) leads to the result

\begin{subequations}\label{Thetaprime} 
\begin{equation}
\hat{\Omega}_{k}^{\theta}=\sum_{N\in\mathbb{Z}}f_{N}\int dq_{\theta'}e^{iN\phi_{k}^{(N)}\left(q_{\theta'}\right)}\left|q_{\theta'}\right\rangle \left\langle q_{\theta'}-N\tau_{\theta}\right|,
\end{equation}
with 
\begin{equation}
\phi_{k}^{(N)}\left(q_{\theta'}\right)=\tau_{\theta}\left(q_{\theta'}-\frac{N\tau_{\theta}}{2}\right)\cot\Delta\theta-\left(2\pi k/d+\frac{q_{\theta}^{\textrm{cen}}\tau_{\theta}}{\sin\Delta\theta}\right),
\end{equation}
\end{subequations}and $\tau_{\theta}=2\pi\sin\Delta\theta/T_{\theta}$.
Note that in the limit $\theta'\rightarrow\theta$, or equivalently
$\Delta\theta\rightarrow0$, the above expressions are not singular,
since $\tau_{\theta}\rightarrow0$ in this limit.

\section{Mutual unbiasedness}

\label{sec:MUB2}

Given a pure state $\left|\Psi\right\rangle $, we define a family
of probabilities 
\begin{equation}
p_{k}^{(\theta)}\left(\Psi\right)=\left\langle \Psi\right|\hat{\Omega}_{k}^{\theta}\left|\Psi\right\rangle ,
\end{equation}
which encodes the results of the PCG measurements in question. Following
\cite{Tasca17a}, measurements labeled by different angles in phase
space, $\theta$ and $\theta'$, are mutually unbiased if for all
$\left|\Psi\right\rangle $ and all $k_{0},l_{0}=0,\ldots,d-1$,

\begin{subequations}\label{MUBs} 
\begin{equation}
p_{k}^{(\theta')}\left(\Psi\right)=\delta_{k_{0}k}\;\Longrightarrow\;p_{l}^{(\theta)}\left(\Psi\right)=d^{-1},\label{MUB1}
\end{equation}
\begin{equation}
p_{l}^{(\theta)}\left(\Psi\right)=\delta_{l_{0}l}\;\Longrightarrow\;p_{k}^{(\theta')}\left(\Psi\right)=d^{-1}.\label{MUB2}
\end{equation}
\end{subequations}

In words, whenever a state is localized according to the measurements
with respect to the phase-space variable specified by the angle $\theta^{\prime}$,
it is evenly spread with respect to measurements corresponding to
the second variable defined by $\theta$. Note that whenever the pairs
of projective measurements are unitarily equivalent, it is sufficient
to consider a single condition, say (\ref{MUB1}). In phase space
this is usually the case since $\hat{q}_{\theta'}=\hat{F}_{\Delta\theta}^{\dagger}\hat{q}_{\theta}\hat{F}_{\Delta\theta}$,
with $\hat{F}_{\Delta\theta}$ being the unitary fractional Fourier
transform operator. However, for the particular projective measurements
considered here, one also needs an extra symmetry. Namely, the presumed
condition on the periods $T_{\theta}$ and $T_{\theta'}$, which is
necessary to fulfill (\ref{MUB1}), must be invariant under the swap
of both periods \cite{Tasca17a}. The last requirement will be verified
below. Here we mainly emphasize that in the discu\textcolor{black}{ssed
scenario the requirement (\ref{MUB2}) follows automatically, provided
that (\ref{MUB1}) is satisfied.}

\textcolor{black}{We are in position to establish the mai}n theoretical
results of this paper. Due to Eq. \eqref{Thetaprime} we can write
\begin{equation}
p_{l}^{(\theta)}\left(\Psi\right)=\sum_{N\in\mathbb{Z}}f_{N}\int dq_{\theta'}e^{iN\phi_{l}^{(N)}\left(q_{\theta'}\right)}\psi^{*}\left(q_{\theta'}\right)\psi\left(q_{\theta'}-N\tau_{\theta}\right),\label{prob}
\end{equation}
with $\psi^{*}\left(q_{\theta'}\right)=\left\langle q_{\theta'}\left|\Psi\right\rangle \right.\!$.
This expression is a direct extension of the formula derived in \cite{Tasca17a}
for the special case of the conjugate pair of position and momentum.
Indeed, if $\theta'=0$ and $\theta=\pi/2$, so that $q_{\theta'}\equiv x$,
we find that $\tau_{\theta}$ is negative and equal to $-2\pi/T_{p}$,
with $T_{p}\equiv T_{\pi/2}$.

Returning to the general case, we immediately conclude from Eq. \eqref{prob}
that if 
\begin{equation}
\frac{T_{\theta'}T_{\theta}}{2\pi}=\frac{d\left|\sin\Delta\theta\right|}{m},\quad m\in\mathbb{N},\quad\textrm{s.t.}\quad\forall_{n=1,\ldots,d-1}\;\frac{m\,n}{d}\notin\mathbb{N},\label{condition}
\end{equation}
then the mutual-unbiasedness condition (\ref{MUBs}) is fulfilled.
In words, $m$ is a natural number \footnote{We assume here that $0$ does not belong to $\mathbb{N}$.}
such that $m\,n/d\notin\mathbb{N}$ for all $n=1,\ldots,d-1$. As
explained in \cite{Tasca17a}, there is no clear pattern followed
by the allowed values of $m$. However, the case $m=1$ stands out
as it is present for a\textcolor{black}{ll values of }$d$. An excluded
case in which $m$ is a multiple of $d$ shall correspond to pairs
of modular variables on the phase space, as this happens for the standard
scenario $\theta'=0$ and $\theta=\pi/2$ \cite{Busch86,Tasca17a}.
Again, both displacements of the origins are absent in (\ref{condition}).

To prove the above statement we observe that under the condition in
question, an autocorrelation term present in (\ref{prob}) simplifies
to $\psi^{*}\left(q_{\theta'}\right)\psi\left(q_{\theta'}-mN\epsilon_{\Delta\theta}T_{\theta'}/d\right)$
with $\epsilon_{\Delta\theta}=\textrm{sign}\left(\sin\Delta\theta\right)$.
Now, if $p_{k}^{(\theta')}\left(\Psi\right)=\delta_{k_{0}k}$ for
some $k_{0}$, then the autocorrelation differs from zero only for
integer values of $mN/d$. The sign $\epsilon_{\Delta\theta}$ plays
no role here. Due to the additional requirement put on $m$, the quantity
$mN/d$ is an integer only when $N/d\in\mathbb{Z}$. In turn, the
prefactor $f_{N}$ defined in (\ref{prefactor}) vanishes for all
$N$ that are multiples of $d$, except the case $N=0$, when it assumes
the value of $1/d$. This proves the desired result, since for $N=0$,
the $q_{\theta'}$ integral in (\ref{prob}) is equal to $1$. As
already mentioned, the condition (\ref{condition}) is invariant with
respect to the swap of both periods, as it only depends on their product.

\section{Mutual unbiasedness for several variables}

\label{sec:MUB3} 

For infinite-dimensional CV systems there are up to three simultaneously
mutually unbiased bases \cite{Weigert08}, given\textendash for example\textendash as
eigenbases of the operators $\hat{x}\equiv\hat{q}_{0}$, $\hat{r}\equiv\hat{q}_{2\pi/3}$
and $\hat{s}\equiv\hat{q}_{4\pi/3}$. Contrarily, in finite-dimensional
quantum mechanics one can find even more MUBs \cite{Durt10}. The
coarse-grained scenario is the discretization of the CV case, situated
somewhat between these two distinct regimes, and thus it is especially
interesting to discover which pattern of mutual unbiasedness will
be reproduced. We start with the general case of the periodic coarse-grained
version of mutually unbiased CV variables, while later focus on the
particular case of $m=1$ {[}see Eq. (\ref{condition}){]}, valid
for all $d$, and prove that, similarly to the usual CVs, only three
PCG measurements can be simultaneously mutually unbiased. In the most
general scenario \textemdash{} those with independent values of $m$
for each pair of variables \textemdash{} there is room for more complex
settings (possibly more than three unbiased measurements). We will
not explore this plethora of possibilities here.

\subsection{Three mutually unbiased CV}

Let $T_{x}$, $T_{r}$ and $T_{s}$ be the periods corresponding to
the mask functions of PCG measurements along $x$, $r$ and $s$.
Since for each pair of the variables $\left|\sin\Delta\theta\right|=\sqrt{3}/2$,
equations (\ref{condition}) for $\theta,\theta'\in\left\{ 0,2\pi/3,4\pi/3\right\} $
become 
\begin{equation}
T_{x}T_{r}=\frac{\sqrt{3}\pi d}{m_{1}},\;\quad T_{x}T_{s}=\frac{\sqrt{3}\pi d}{m_{2}},\;\quad T_{r}T_{s}=\frac{\sqrt{3}\pi d}{m_{3}},
\end{equation}
where $m_{1},m_{2},m_{3}$ are natural numbers satisfying the relevant
constraints in Eq. (\ref{condition}). The general solution is found
to be: 
\begin{equation}
T_{x}=\sqrt{\sqrt{3}\pi d\frac{m_{3}}{m_{1}m_{2}}},\quad T_{r}=\frac{m_{2}}{m_{3}}T_{x},\quad T_{s}=\frac{m_{1}}{m_{3}}T_{x}.\label{solutions}
\end{equation}

Clearly, for the special case $m_{1}=m_{2}=m_{3}=1$, all the periods
assume the fixed, dimension-dependent value of $\sqrt{\sqrt{3}\pi d}$.
Contrary to the case of the mutually unbiased pair, in which one of
the periods is not fixed but serves as a reference length \cite{Tasca17a},
for the mutually unbiased triple all the periods are fixed, up to
the freedom offered by the natural numbers $m$. This situation is
very similar to saturation of variance-based uncertainty relations.
For two variables, all pure Gaussian states sat\textcolor{black}{urate
a version of the Heisenberg uncertainty relation, regardless of the
variance of the state. For three CV variables, the product of three
variances is bounded by $1/8$ \cite{Weigert08,Paul16,Dodonov18},
and the inequality $\left(\Delta x\right)^{2}\left(\Delta r\right)^{2}\left(\Delta s\right)^{2}\geq1/8$
is saturated only when all the invo}lved variances are equal to $1/2$.

\subsection{Maximum Number of MUBs in the case $m=1$}

\textcolor{black}{Here we show that for the periodic coarse graining
considered and with $m=1$, a maximum number of three MUBs is possible.
We }proceed by searching for an arbitrary number of $K$ coarse-grained,
periodic MUBs, specified by the angles $\theta_{1},\ldots,\theta_{K}$
and the periods $T_{1},\ldots,T_{K}$. Without loss of generality
we assume that $\theta_{j}>\theta_{i}$ whenever $j>i$, and that
$0\leq\theta_{i}<2\pi$ for all $i=1,\ldots,K$.

We would like to check what number $K$ of PCG measurements $\left\{ \hat{\Omega}_{k}^{\theta_{j}}\right\} $
can simultaneously be mutually unbiased with the same value $m=1$,
i.e. the conditions

\begin{equation}
\frac{T_{i}T_{j}}{2\pi}=d\left|\sin\left(\theta_{i}-\theta_{j}\right)\right|,
\end{equation}
are all satisfied. Using $i=1$ we get ($j\geq2$): 
\begin{equation}
\frac{T_{j}}{2\pi}=\frac{d}{T_{1}}\left|\sin\left(\theta_{1}-\theta_{j}\right)\right|,
\end{equation}
so that, after minor trigonometric simplification, we are left with
$\left(K-1\right)\left(K-2\right)/2$ consistency conditions ($j>i\geq2$):
\begin{equation}
\left|\cot\left(\theta_{j}-\theta_{1}\right)-\cot\left(\theta_{i}-\theta_{1}\right)\right|=\frac{2\pi d}{T_{1}^{2}}.
\end{equation}

It is clear that if $K=3$ we have a single condition for $i=2$ and
$j=3$, which in turn can trivially be solved for $T_{1}$. In that
way one immediately obtains a generalized ($\theta$-dependent) variant
of the solutions (\ref{solutions}).

Analyzing the case $K=4$, it is convenient to make a simplifying
assumption which does not spoil the generality of the argument: let
us assume that $\theta_{1}=0$, i.e. our first member is the periodic
coarse graining of the position variable. In this case we are left
with three consistency conditions, which, after solving one of them
with respect to $T_{1}$, lead to two equations:

\begin{subequations} 
\begin{equation}
\left|\cot\theta_{3}-\cot\theta_{2}\right|=\left|\cot\theta_{4}-\cot\theta_{2}\right|,
\end{equation}
\begin{equation}
\left|\cot\theta_{4}-\cot\theta_{3}\right|=\left|\cot\theta_{4}-\cot\theta_{2}\right|.
\end{equation}
\end{subequations}

All three angles $\theta_{2},\theta_{3},\theta_{4}$ can neither be
$0$ nor $\pi$ as in such a case some of the directions would reproduce
the position variable (or $-x$). As a result, no infinities occur
in the above conditions. By a similar argument we exclude $\cot\theta_{2}=0$,
since it implies $\cot\theta_{3}=0=\cot\theta_{4}$, and all three
directions would need to correspond to $\pm p$. Since $\cot\theta_{2}\neq0$
we can introduce auxiliary variables $\zeta_{3}=\cot\theta_{3}/\cot\theta_{2}$
and $\zeta_{4}=\cot\theta_{4}/\cot\theta_{2}$, and rewrite the consistency
conditions in the form 
\begin{equation}
\left|\zeta_{3}-1\right|=\left|\zeta_{4}-1\right|=\left|\zeta_{4}-\zeta_{3}\right|.
\end{equation}

It is easy to verify that the only solution to these conditions is
given by $\zeta_{3}=1=\zeta_{4}$. As a result we need to find three
angles from the range $\left]0,2\pi\right[$ for which their cotangent
would assume the same value. Since in the desired range the cotangent
function assumes every real, finite value exactly twice, it is impossible
to fit four coarse-grained MUBs (with $m=1$) into the phase space.
This argument can be applied to any number $K>3$ directions in phase
space. Thus, for the PCG considered, there are at most 3 mutually
unbiased measurements.

\section{Experiment}

\label{sec:exp}

\subsection{Three mutually unbiased PCG measurements}

To demonstrate mutually unbiased PCG measurements of the phase-space
triple $x\;(\theta=0)$, $r\;(\theta=2\pi/3)$ and $s\;(\theta=4\pi/3)$,
we performed an optics experiment exploring the transverse spatial
variables of a paraxial laser beam. Using systems of converging lenses
and spatial light modulators (SLMs), we prepared the eigenstate of
the PCG measurement operator $\hat{\Omega}_{k}^{\theta}$, described
by Eq.~\eqref{Projectors} with Eq.~\eqref{eq:mask-definition}.
In the sequence, we used an equivalent apparatus to perform the measurement
described by another o\textcolor{black}{perator $\hat{\Omega}_{k'}^{\theta'}$
on the prepared state. T}his procedure was followed for every combination
of (different) $\theta,\theta'\in\{0,\nicefrac{2\pi}{3},\nicefrac{4\pi}{3}\}$,
and every possible value of $k$ and $k'$ for measurement dimensionality
$d$ from 2 up to 10. The PCG measurements $\{\hat{\Omega}_{k}^{\theta}\}$
and $\{\hat{\Omega}_{k'}^{\theta'}\}$ can be considered unbiased
if the resulting conditional probability distributions satisfy $p(k'|k)=\nicefrac{1}{d}\;\forall\;k,k'$. 

\begin{figure}
\includegraphics[width=8.5cm]{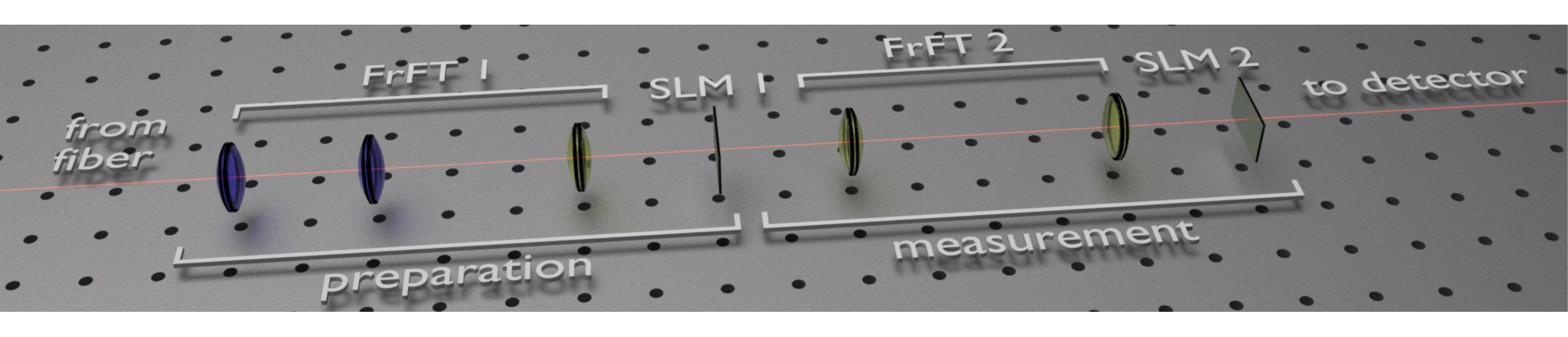}

\caption{\label{fig:experimental setup}Sketch of the experimental setup. The
actual spatial light modulators (SLM) used were reflective. The fractional
Fourier Transforms (FrFT) are described in the text.}

\end{figure}

Fig. \ref{fig:experimental setup} shows a simplified scheme of our
experimental setup. The output of an attenuated Thorlabs' CPS180 635-nm
diode laser was first coupled in and out of a single-mode fiber. At
the output of the fiber the beam was well-collimated so that its transverse
field distribution was given by $\exp[-x^{2}/(4\sigma^{2})]$, where
$\sigma=875\;\mathrm{\mu m}$. Though our experiment used a laser
beam, the same physics applies to single photons with equivalent spatial
and spectral properties.

The preparation of the eigenstate of operator $\hat{\Omega}_{k}^{\theta}$
was done in two steps. First, the beam was subjected to a fractional
Fourier transform (FrFT) by a set of converging lenses connecting
a reference plane to the plane of SLM 1. In this way, the chosen phase-space
direction, defined by $\theta$, was mapped onto the phase-space $x$
axis on the physical position of SLM 1, which then performed the operation
described by $\hat{\Omega}_{k}^{0}$. This procedure allows us to
prepare eigenstates of the operator $\hat{\Omega}_{k}^{\theta}$.

\begin{figure}
\begin{centering}
\includegraphics[width=8cm]{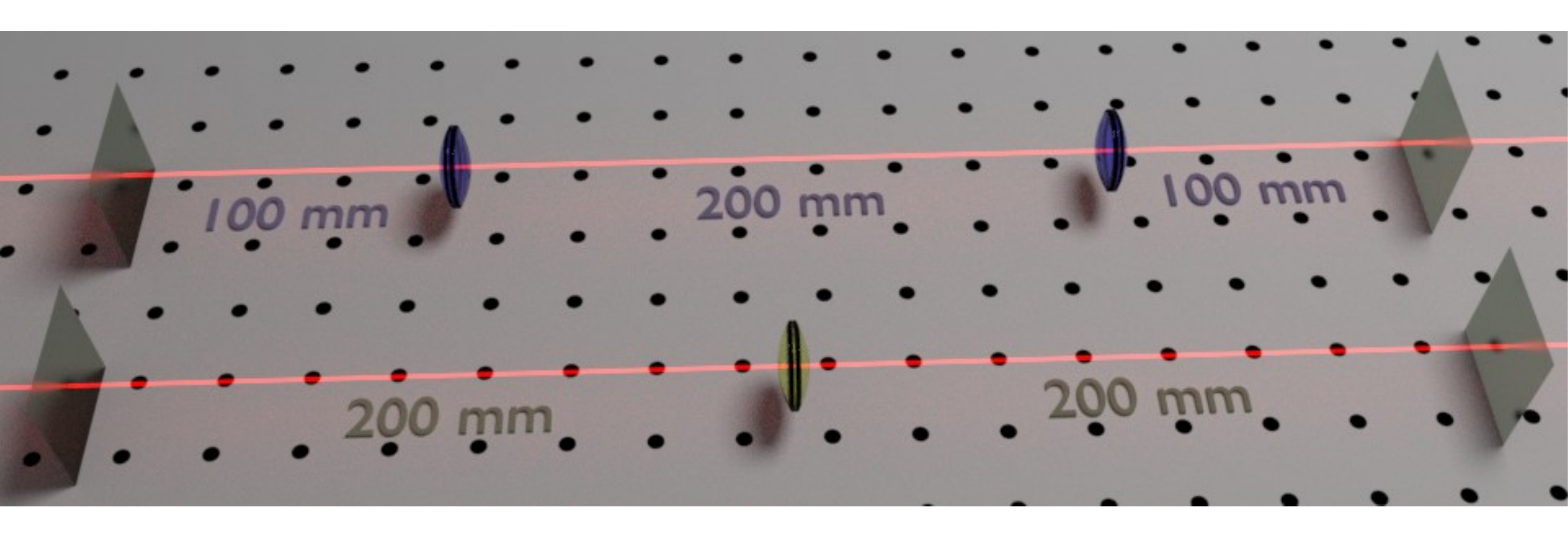} 
\par\end{centering}
\caption{\label{fig:Lens-systems-used}Lens systems used to perform the FrFTs.
On the top, two confocal 100-mm focal-length lenses are used to perform
a \textgreek{p} rotation. On the bottom, a single 400-mm lens is used
to perform a \textgreek{p}/3 rotation. The planes represent the object
and image planes of the systems. Sequential combinations of these
systems were used to produce the desired phase-space rotations.}
\end{figure}

A single lens of focal length $f$, placed symmetrically at a distance
$z$ from the input and output planes, performs a FrFT characterized
by the phase-space rotation angle $\theta$ (with proper dimensionalization
discussed below), where \cite{Lohmann93}
\begin{equation}
z=2f\sin(\theta^{2}/2).\label{eq:FrFT_relation}
\end{equation}

In order to achieve all of the phase-space rotations desired, we combined
two different systems of lenses. On one hand, a pair of confocal 100-mm
focal-length lenses together performed a \textgreek{p} rotation (see
Fig.~\ref{fig:Lens-systems-used}). On the other hand, a single 400-mm
focal-length lens, placed 200 mm away from both its input and output
planes performed a \textgreek{p}/3 rotation \cite{Tasca11} (see also
Fig.~\ref{fig:Lens-systems-used}). Therefore, two sequential 400-mm
lenses performed a 2\textgreek{p}/3 rotation, a pair of 100-mm lenses
followed by a 400-mm lens performed a 4\textgreek{p}/3 rotation, and
two sequential pairs of 100-mm lenses were used to perform a 2\textgreek{p}
rotation.

To adequately interpret the action of a single lens as a FrFT and
therefore as a rotation in phase space, it is necessary to use dimensionless
variables. This can be done by using the scaling factor \cite{Tasca11}
\begin{equation}
\delta=\sqrt{\lambda f\sin\theta/2\pi},\label{scaling factor}
\end{equation}
where $\lambda$ is the laser beam's wavelength, $f$ is the lens'
focal length and $\theta$ is the corresponding phase-space rotation
angle. Choosing $f$ = 400 mm and $\theta=\pi/3$ we were able to
use the same scaling factor for all variables. This means the dimensionless
variables are $x=x'/\delta$, $r=r'/\delta$ and $s=s'/\delta$, where
the primed letters represent the measured transverse spatial variables
(with dimension of length).

\begin{table}
\begin{tabular}{|c|c|c|c|}
\hline 
$d$  & $T$ (\textgreek{m}m)  & $T/l$  & $T_{exp}$ (\textgreek{m}m)\tabularnewline
\hline 
\hline 
2  & 617.3  & 77.2  & 616\tabularnewline
\hline 
3  & 756.0  & 94.5  & 752\tabularnewline
\hline 
4  & 872.9  & 109.1  & 872\tabularnewline
\hline 
5  & 976.0  & 122.0  & 976\tabularnewline
\hline 
6  & 1069.1  & 133.6  & 1072\tabularnewline
\hline 
7  & 1154.8  & 144.3  & 1152\tabularnewline
\hline 
8  & 1234.5  & 154.3  & 1232\tabularnewline
\hline 
9  & 1309.4  & 163.7  & 1312\tabularnewline
\hline 
10  & 1380.2  & 172.5  & 1384\tabularnewline
\hline 
\end{tabular}

\caption{\label{tab:periods}Ideal mask periods ($T$), as calculated from
the theory, and their ratio to the SLM pixel length ($T/l$), as a
function of the dimension ($d$). The actual experimental periods
used ($T_{exp}$) correspond to the nearest possible integer number
of pixels.}
\end{table}

To perform the operations described by $\hat{\Omega}_{k}^{0}$, we
used reflective Holoeye Pluto phase-only SLMs, which generated the
desired mask functions aligned with the horizontal (lab table). All
the mask periods were given by Eq.~\eqref{solutions}, with $m_{1}=m_{2}=m_{3}=1$.
Combining equations Eqs. \eqref{solutions} and \eqref{scaling factor},
we can calculate the periods for each dimension. Considering that
we were constrained to using periods that were integer multiples of
the SLM pixel length ($l$ = 8 \textgreek{m}m), the actual periods
used were approximations of the theoretical ones, as we can see in
Table \ref{tab:periods}.

After the prepare and measure procedures, the beam size was then reduced
by a factor of $\approx10$ (a magnification of $\approx0.1$) by
an imaging system consisting of two confocal lenses of focal lengths
250 mm and 25.4 mm, and coupled into a 300-\textgreek{m}m core multi-mode
fiber using a 10X Olympus plane achromat objective. The multi-mode
fiber was then connected to one channel of a Perkin Elmer SPCM-AQ4C
single-photon avalanche photodiode detector. The number of photons
arriving at the detector in a 0.1 s time interval was registered.

\begin{figure}
\centering{}%
\begin{tabular}{lll}
\includegraphics[width=8cm]{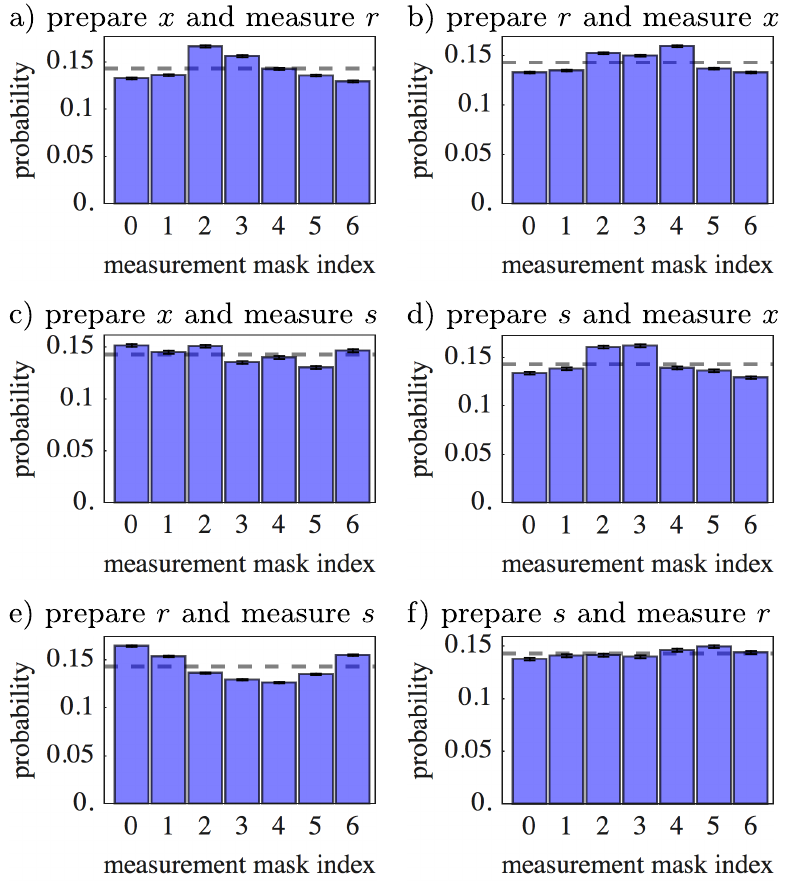}
\end{tabular}

\caption{\label{fig:7-dim-results}Probability distributions obtained for a
periodic coarse graining with dimension $d=7$, when using preparation
mask $k=2$. Each of the graphs corresponds to one of the possible
combinations of the prepare and measure phase-space directions $x\left(\theta=0\right)$,
$r\left(\theta=2\pi/3\right)$ and $s\left(\theta=4\pi/3\right)$.
The dashed lines correspond to the theoretical predictions. The error
bars correspond to the standard deviations derived from the laser
poissonian count statistics.}
\end{figure}

For each preparation, all the measurement outcomes were normalized
with respect to the sum of all outcomes (for the preparation at hand).
The normalized values were then interpreted as the conditional probabilities
$p(k'|k)$ of obtaining each measurement result $k'$ given the corresponding
preparation $k$. Fig.~\ref{fig:7-dim-results} shows an example
of the probability distributions obtained for $d=7$, where the prepared
mask was $k=2$. One can see qualitatively that the measurement outcomes
for all pairs of variables are approximately uniform, displaying three-fold
mutual unbiasedness for these periodic coarse-grained measurements.

\begin{figure}
\includegraphics[width=8cm]{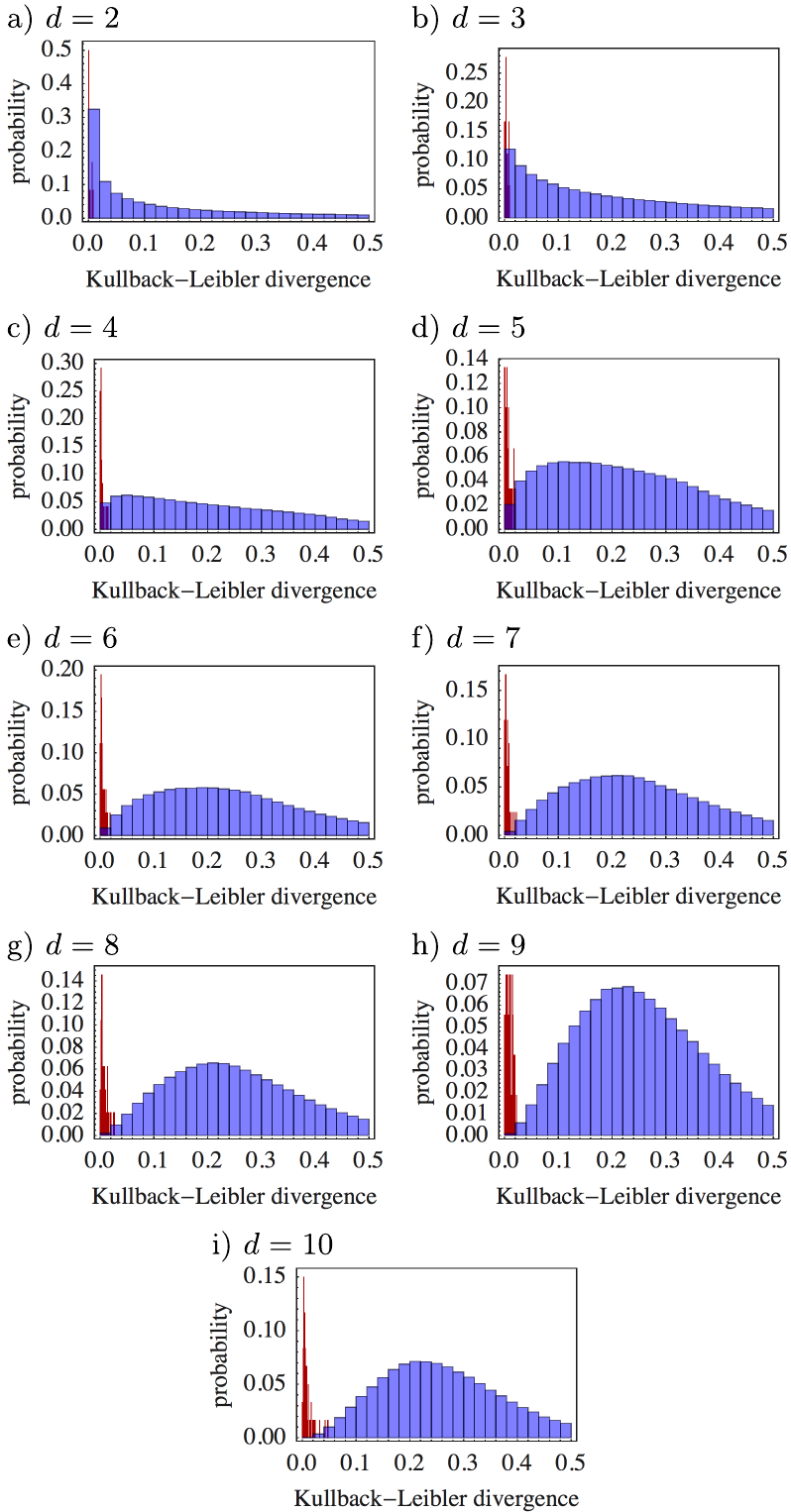}
\caption{\label{fig:histograms}Histograms of the Kullback-Leibler divergences.
The blue bars represent the $10^{6}$ simulated random distributions,
while the red bars represent the experimental data. The horizontal
axes are the KL divergences and the vertical axes are the probabilities
of occurrence.}
\end{figure}

As a figure of merit to check the measurements' unbiasedness, we opted
to use the Kullback-Leibler (KL) divergence, also known as the relative
entropy\textcolor{black}{:}
\begin{equation}
D(P||Q)=\sum_{i=0}^{d-1}P_{i}\log\left(\frac{P_{i}}{Q_{i}}\right),
\end{equation}
where $P$ is the measured probability distribution, $Q$ is the target
uniform distribution, such that $Q_{i}=1/d$ for all $Q_{i}$, and
we use the base-2 logarithm. A perfect match between generated and
target uniform distribution means zero divergence, while a maximal
divergence of $D=\log(d)$ is achieved for a perfectly localized distribution.
Using this quantity, we compared (for each dimension) the set of experimental
probability distributions to a sample of $10^{6}$ simulated random
probability distributions. Each of Figs. \ref{fig:histograms} (one
for each \textcolor{black}{value of $d$}) shows the histogram of
KL divergences for the sample of simulated probability distributions
(blue bars), together with the histogram for the divergences calculated
from the experimentally obtained probability distributions (red bars).
We can clearly see from the graphs that the experimental probability
distributions produce KL divergences much lower than what would be
expected from random chance.

\begin{table}
\centering{}%
\begin{tabular}{|c|c|}
\hline 
$d$  & $P$\tabularnewline
\hline 
\hline 
2  & $80\pm1$\tabularnewline
\hline 
3  & $94.4\pm0.4$\tabularnewline
\hline 
4  & $97.2\pm0.2$\tabularnewline
\hline 
5  & $98.2\pm0.2$\tabularnewline
\hline 
6  & $99.3\pm0.1$\tabularnewline
\hline 
7  & $99.50\pm0.04$\tabularnewline
\hline 
8  & $99.65\pm0.03$\tabularnewline
\hline 
9  & $99.89\pm0.02$\tabularnewline
\hline 
10  & $99.33\pm0.04$\tabularnewline
\hline 
\end{tabular}\caption{\label{tab:KL-divergences}Percentage ($P$) for each dimension ($d)$
of simulated probability distributions that produce KL divergences
greater than the greatest value achieved for the experimental distributions.
The errors listed correspond to the standard deviations.}
\end{table}

To have a better quantitative view of these results, we calculated
(again, f\textcolor{black}{or each $d$) the} percentage of simulated
probability distributions that produce KL divergences greater than
the greatest value achieved for the experimental distributions (see
Table \ref{tab:KL-divergences}). If we allow the extrapolation of
these results to the set of every (infinite) $d$-dimensional probability
distribution, this means that each experimental distribution is closer
to the uniform distribution, as per the KL divergence, than at least
the given percentage of every possible $d$-dimensional probability
distribution. These results show that we were able to obtain probability
distributions very close to the uniform distribution, and that their
proximity was not solely a matter of random chance, demonstrating
that the three measurements defined are indeed mutually unbiased.

\subsection{Unbiasedness for two \textcolor{black}{arbitrary} directions in phase
space}

\begin{figure}
\includegraphics[width=8cm]{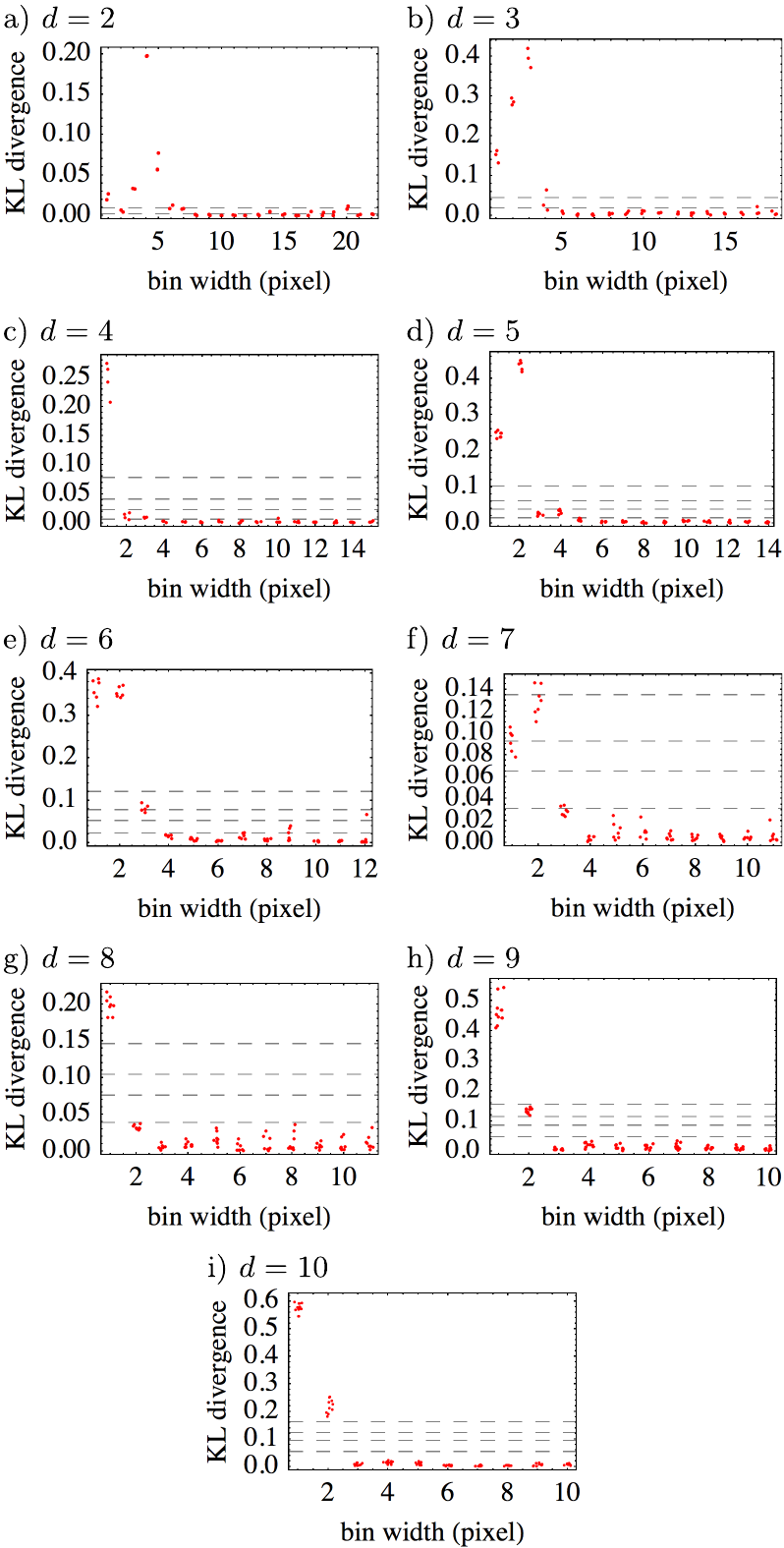}
\caption{\label{fig:KL_divergences_xDelta}The Kullback-Leibler divergences
of the distributions obtained by preparing an eigenstate of $\hat{\Omega}_{k}^{0}$
and performing the measurements described by $\left\{ \hat{\Omega}_{k'}^{23\protect\textdegree}\right\} $.
Each figure summarizes the results for one dimension. The horizontal
axes are the bin widths in terms of number of pixels. The vertical
axes are the KL divergences from the uniform distributions. The top
dashed line in each figure identifies the KL divergence value above
which lie $80\%$ of the probability distributions simulated in the
previous section. Analogously, the following lines from top to bottom
determine the limits for $90\%$, $95\%$ and $99\%$ of values respectively
(note that in the first two figures, only the lines corresponding
to the $80\%$ and $90\%$ limits are seen). The error bars have been
omitted for being about the size of the points.}
\end{figure}

Any two PCG measurements in distinct directions in phase space are
mutually unbiased, given that Eq.~\eqref{condition} is satisfied.
With that in mind, we carried out experiments in which we prepared
eigenstates of $\hat{\Omega}_{k}^{0}$, for every $k=0,...,d-1$,
and then performed measurements described by the operators $\left\{ \hat{\Omega}_{k'}^{\alpha}\right\} $,
with the smallest value for the a\textcolor{black}{ngle $\alpha$
that we could} achieve in our experimental setup. Using $z=200\,\mathrm{mm}$
and $f=250\,\mathrm{mm}$, Eq.~\eqref{eq:FrFT_relation} gives $\theta\approx78.5\textdegree$.
Two such optical systems in a row implement a phase-space rotation
of $157\textdegree$. Following this operation by a reflection in
real space, which is equivalent to the transformation $x\rightarrow-x$
and is automatically fulfilled due to the reflective character of
the SLMs, we achieved the rotation angle of $\alpha\approx23\textdegree$.
With this operation, we performed the same procedure described in
the previous section, but repeating the measurements for several bin
widths. Each one of Figs.~\ref{fig:KL_divergences_xDelta} shows
the KL divergences from the uniform distribution obtained for these
measurements, one for e\textcolor{black}{ach $d$. O}n the horizontal
axes are the bin widths, in pixels, used for the preparation masks.
The bin width of the measurement masks can be obtained from Eq.~\eqref{condition}.
We can see that for the smallest bin widths, the data te\textcolor{black}{nd
to not be as reliable. Nevertheless, for larger bin widths the KL
divergences are clearly near zero.}

\textcolor{black}{Again, we can see that, even for a relatively small
angle in phase-space between the variables, w}e obtain probability
distributions very close to the uniform distribution, showing the
unbiasedness of the measurements.

\section{Discussion}

\textcolor{black}{Periodic coarse graining can be used to recover
mutual unbiasedness, a property that is not present in physical realization
of the standard coarse-grained versions of phase-space observables.
We have shown that, for a periodic coarse graining given by mask functions
defined in Eq. \eqref{eq:mask-definition}, and in the particular
case of $m_{1}=m_{2}=m_{3}$ {[}see Eq. \eqref{solutions}{]}, up
to three mutually unbiased bases can be defined. The open question
remains as to whether one can define more mutually unbiased measurements
for other cases, and is an interesting topic to be investigated.}

\textcolor{black}{An optics experiment was performed to demonstrate
these results. The optical fractional Fourier transform was used to
prepare and measure perio}dic coarse-grained versions of a symmetric
phase space triple of operators. Very good agreement was found between
theory and experiment for all measurement dimensionality tested in
our experiment (from $2$ to \textcolor{black}{$10$). In addition,
we experimentally tested mutual unbiasedness for periodic coarse graining
of measurements that correspond to phase-space variables that are
separated by only a $23\textdegree$ rotation in phase space.}

Our theoretical and experimental results contribute to the further
understanding of the relation between continuous and discrete quantum
mechanics, and could prove useful in the discover of new uncertainty
relations, and in adapting quantum information protocols to continuous
variable systems.

\label{sec:conc} \begin{acknowledgements} The authors acknowledge
financial support from the Brazilian funding agencies CNPq, CAPES
and FAPERJ, and the National Institute of Science and Technology -
Quantum Information. \L .R. acknowledges financial support by grant
number 2014/13/D/ST2/01886 of the National Science Center, Poland.
\end{acknowledgements}

\appendix

\section{Derivation of Eq. \eqref{Thetaprime}}

\label{ApA} We start the derivation by employing the Fourier series
expansion (\ref{FourierSeries}) of the periodic mask function 
\begin{equation}
\hat{\Omega}_{k}^{\theta}=\sum_{N\in\mathbb{Z}}f_{N}e^{iN\varphi_{k}}\int dq_{\theta}e^{i\frac{N\tau_{\theta}q_{\theta}}{\sin\Delta\theta}}\left|q_{\theta}\right\rangle \left\langle q_{\theta}\right|,
\end{equation}
with $\varphi_{k}=-2\pi k/d-q_{\theta}^{\textrm{cen}}\tau_{\theta}/\sin\Delta\theta$.
Using the completeness relation $1=\int dq_{\theta'}\left|q_{\theta'}\right\rangle \left\langle q_{\theta'}\right|$,
we can change the basis from $\left|q_{\theta}\right\rangle $ to
its rotated counterpart $\left|q_{\theta'}\right\rangle $

\begin{equation}
\hat{\Omega}_{k}^{\theta}=\sum_{N\in\mathbb{Z}}f_{N}e^{iN\varphi_{k}}\int dq_{\theta'}\int d\tilde{q}_{\theta'}Q\left(q_{\theta'},\tilde{q}_{\theta'}\right)\left|q_{\theta'}\right\rangle \left\langle \tilde{q}_{\theta'}\right|,\label{withQ}
\end{equation}
where 
\begin{equation}
Q\left(q_{\theta'},\tilde{q}_{\theta'}\right)=\int dq_{\theta}e^{i\frac{N\tau_{\theta}q_{\theta}}{\sin\Delta\theta}}\mathcal{F}\left(q_{\theta'},q_{\theta}\right)\mathcal{F}\left(q_{\theta},\tilde{q}_{\theta'}\right).
\end{equation}
The variable $\tilde{q}_{\theta'}$ has the same physical meaning
as $q_{\theta'}$ and the tilde on top of it has only been introduced
to distinguish both integration variables.

An explicit form of the above kernel is 
\begin{equation}
Q\left(q_{\theta'},\tilde{q}_{\theta'}\right)=\frac{e^{i\frac{\cot\Delta\theta}{2}\left(q_{\theta'}^{2}-\tilde{q}_{\theta'}^{2}\right)}}{2\pi\left|\sin\Delta\theta\right|}\int dq_{\theta}e^{i\frac{q_{\theta}}{\sin\Delta\theta}\left(N\tau_{\theta}+\tilde{q}_{\theta'}-q_{\theta'}\right)}.
\end{equation}

One can immediately perform the integration with respect to $q_{\theta}$
as it leads to the Dirac delta function 
\begin{equation}
Q\left(q_{\theta'},\tilde{q}_{\theta'}\right)=e^{i\frac{\cot\Delta\theta}{2}\left(q_{\theta'}^{2}-\tilde{q}_{\theta'}^{2}\right)}\delta\left(N\tau_{\theta}+\tilde{q}_{\theta'}-q_{\theta'}\right).
\end{equation}

In the last step, one needs to plug the above expression into Eq.
\ref{withQ}, perform the integration with respect to $\tilde{q}_{\theta'}$
and simplify accordingly. 


\end{document}